%% file: cncurr.acm.tex
\documentclass{acmtrans2m}

\usepackage{amsmath}
\usepackage{amssymb}
\usepackage{amsmath}
\usepackage{verbatim}
\usepackage{epsfig}

\input{commands.acm.tex}

\title{Complexity of Data Dependence problems for  Program  Schemas with Concurrency}

\author{SEBASTIAN DANI{C}I{C}\\
Department of Computing, 	  	 
Goldsmiths, 	  	 
University of London, 	  	 
London, SE14 6NW, 
UK \\    
ROBERT M. HIERONS\\
Department of Information Systems and Computing, Brunel University, Uxbridge, Middlesex, UB8 3PH
\and
MICHAEL R. LAURENCE\\ 
Department of Computer Science,
Regent Court, 211 Portobello, Sheffield, S1 4DP, UK }

\begin{abstract} \noindent The problem of deciding whether one point in a program is data dependent upon another is fundamental to program analysis and has been widely studied. In this paper we consider this problem at the abstraction level of {\it program schemas} in which computations occur in the Herbrand domain of terms and predicate symbols, which represent arbitrary predicate functions, are allowed. Given a vertex $l$ in the flowchart of a schema $S$ having only equality (variable copying)  \asss,\ and variables $v,w$,  we show that it is PSPACE-hard to decide whether \txs\ an execution of a program defined by $S$ in which $v$  holds the initial value of $w$ at at least one occurrence of $l$ on the path of execution, with membership in PSPACE holding provided there is a constant upper bound on the arity of any predicate in $S$.
 We also consider the `dual' problem in which $v$ is required to hold the initial value of $w$ at every occurrence of $l$, for which the analogous results hold. Additionally, the former problem for programs with non-deterministic branching (in effect, free schemas) in which \asss\ with functions are allowed is proved to be polynomial-time decidable provided a constant upper bound is placed upon the number of occurrences of the concurrency operator in the schemas being considered. This result is promising since many concurrent systems have a relatively small number of threads (concurrent processes), especially when compared with the number of statements they have.
  \end{abstract}
  
  \category{D.3.3}{Programming Languages}{Language Constructs --- 
control structures}
\category{D.3.4}{Programming Languages}{Processors --- 
compilers, optimisation}
\terms{Algorithms, Languages, Design}
\keywords{Program analysis, data dependence, program schemas, 
free schemas, Herbrand domain}

\begin{document}

\begin{bottomstuff}
This work is partly funded by UK Engineering and Physical Sciences Research Council grant EP/E002919/1.
\end{bottomstuff}

\maketitle

\renewcommand{\labelenumi}{{\rm(\arabic{enumi})}}

\def\gr{\mathit{Flowchart}}
\def\edgtyp{\mathit{edgeType}}
\def\labs{\mathit{Labels}}

%\setlength{\baselineskip}{2\baselineskip}

%\fontsize{15.5}{20}\selectfont

%\noindent

\section{Introduction}   \label{intro.sect}

A schema represents the 
statement structure of a program by replacing real functions and predicates by symbols representing them.  
A schema, ${S}$, thus defines a whole class  of programs which  all have the same
structure. 
Each   program  can be obtained from ${S}$ via  a domain $D$ and an {\it interpretation} $i$ which defines a function $f^i: D^n \to D$ for each function symbol $f$ of arity $n$, and a predicate function $p^i: D^m \to \tf$ for each predicate symbol $p$ of arity $m$.
As an example, Figure \ref{S.fig} gives a schema $S$, and the program $P$ of Figure \ref{P.fig} is defined from $S$ by 
interpreting the function symbols $f,g,h$ and the predicate symbol $p$ as given by $P$, with $D$ being the set of integers. 
 The subject of schema theory is connected with that of program  transformation and  was originally motivated by the wish to compile programs     effectively\cite{greibach:theory}. Many results on schema equivalence  \cite{mletal:lfl,lfl:tr,mletal:cfl,sabelfeld:algorithm,luckham:formalised} and on applying schema formulation to program slicing \cite{laurence:flfl,sdetal:lpr} have been published. 
%Since  program slicing algorithms do not normally take into account the meanings of the functions and predicates of a program, a schema encodes all the information about any program which it defines that is available to slicing  algorithms. 

\begin{figure}[t]
\begin{center}
$
\begin{array}{llll}
u\as h(); \\
\si  p(u) && \then 
& v\as f(u); \\
&& \els 
&  v\as g();
 \end{array}
 $ 
\end{center}
\caption{Schema $S$} \label{S.fig}
\end{figure}

\begin{figure}[b]

\begin{center}
$
\begin{array}{llll}
& \\
u\as 1; \\
\si  u> 1 && \then 
& v\as u+1; \\
&& \els 
&  v\as 2;
 \end{array}
 $
\end{center}
\caption{Program $P$} \label{P.fig}
\end{figure}

In this paper we are concerned with establishing complexity bounds for data dependence problems defined on schemas. 
 We only consider schema interpretations over the Herbrand domain of terms in the variables and function symbols. 
We consider the problem of deciding the following two properties, defined using a schema $S$, a variable $v$, a variable or function symbol $f$ and a vertex $l$ in the flowchart of $S$.

\begin{itemize}
\item
(Existential data dependence.) If there is an executable path through $S$ that ends at $l$ at which point the term defined by $v$ contains the symbol $f$, then $\sed(f,v,l)$ is said to hold.
\item
(Universal data dependence.) If, for all executable paths through $S$, the term defined by $v$ contains the symbol $f$ whenever $l$ is reached, then $\sad(f,v,l)$ is said to hold.
\end{itemize}

If $S$ belongs to  the class of schemas in which all
 \asss\  are equality \asss\ (that is, \asss\  of the form $v \as w;$ in which the value held by a variable  $w$ is copied to $v$), we prove the following.

\begin{itemize}
\item 
The problems defined by these properties are both PSPACE-hard, even when $S$ is further required to belong to the class of schemas in which  no concurrency constructs are allowed and only two while loops are permitted in $S$, one of which lies in the body of the other, and no predicate symbol occurs more than once. 
\item
If $S$ is required to contain no loops or concurrency constructs, and each of its predicate symbols has zero arity,   then $\sed(f,v,l)$ is NP-hard, and $\sad(f,v,l)$ is co-NP-hard.
 \item
Both problems lie  in PSPACE  provided there is a constant upper bound on the arity of any predicate in $S$.
\end{itemize}

Additionally, we consider the existential data dependence problem in the case where \asss\ having function symbols are allowed, but where all schemas are free (that is, all paths are executable) and hence all branching is, in effect, non-deterministic. One possible application of data dependence on a function symbol $f$ would be in the case where 
  $f$  corresponds to a call to a function or method that we are altering; we might then want to decide whether this change can propagate through to the value of a particular variable at a particular point. For the class of free schemas, we prove the following.

\begin{itemize}
\item 
 Deciding existential data dependence is  shown to be 
PSPACE-complete, owing to a reduction from the finite intersection problem for deterministic finite state automata.
\item
Under the further condition that  a  constant upper bound is placed upon the number of occurrences of the concurrency operator in the schemas being considered, existential data dependence then becomes decidable in polynomial time. 
\end{itemize}

To the authors' knowledge, neither problem has been  previously considered for arbitrary schemas. Both problems have been studied for programs of various types. In \cite{optimal.slic.parallel.progs.mueller-olm.seidl}, it is proved that 
deciding existential data dependence (expressed in the paper as a slicing problem) is PSPACE-complete for programs having concurrency constructs, but only non-deterministic branching. M\"uller-Olm et al. have also considered a generalisation of our universal data dependence problem \shortcite{mueller-olm:check.herbrand.eq.beyond,DBLP:conf/esop/Muller-OlmSS05}, defined by testing for equality between two terms  at particular program points, but their programs use term inequality guards on edges in flowcharts, and apart from this restriction, their programs are non-deterministic. 
In \cite{olm:complex.constant.propag}, an extensive classification of the complexity of deciding both our problems is given, but branching is non-deterministic and the domain is that of the integers in every case. 

%\begin{comment}
Schemas represent a significantly closer approximation to real-life programs than  purely non-deterministic programs, even when these are very simple. To demonstrate this, consider the schema $S$ in Figure \ref{prog.sch.more.precise.fig}, in which $x,y$ and $z$ are distinct variables.

\begin{figure}
\begin{center}
$
\begin{array}{lll}  & x \as f(); \\
                   & \whi p(z) & x \as g();  \\
                   &l:  & y \as x; \\
 \end{array}
$
\end{center} \caption{A schema demonstrating the greater precision obtainable by considering data dependence problems defined on program schemas compared with non-deterministic programs. } \label{prog.sch.more.precise.fig}
\end{figure}

 Clearly $\sed(g,y,l)$ does not hold, since execution cannot enter the while loop in $S$ and subsequently leave it, whereas if the while loop is replaced by the line $ \lop  x \as g();$ to give a non-deterministic schema $T$, then 
$\sed[T](g,y,l)$ holds. This example motivates extending the study of data dependence problems to schemas, since the gain in precision may be considerable. Another justification for considering program schemas is given by the fact that they have precisely the same level of abstraction as is usually assumed in  program slicing.

As an example of the use of establishing universal data dependence, consider a program which calculates the cost of a purchase - we would expect the overall price to depend always on the costs and amounts of the item(s) purchased. If this fails, then the program clearly contains a fault. 

The complexity results for existential dependence are more promising than they might initially appear. This is because many concurrent systems have only relatively few threads even if they are quite large (in terms of lines of code). The results also suggest that it should be easier to `scale' data dependence algorithms to large programs/schemas with only a few threads than to smaller programs/schemas with many threads. For schemas and programs that might not be free, data dependence calculated on the assumption that freeness holds provides a conservative abstraction of the actual data dependence. As a result, if existential data dependence does not hold under the freeness assumption then we know it does not hold even if the program or schema under consideration is not free. This is important in areas such as security where we wish to show that the value of one variable $x$, whose value is accessible, cannot depend on the value of another variable $y$ whose value should be kept secret.

\section{Basic  definitions  for schemas} \label{basic.sect}

Throughout this paper,  $\f$, $\p$,  $\var$ and $\labels$ denote 
%Without loss of generality one may assume that 
fixed infinite sets of \emph{function symbols},  
\emph{predicate symbols}, \emph{variables} and \emph{labels} respectively. 
%Each element of $ {\f} \cup {\p}$ 
We assume  a  function
$$\arity: \f \cup\p \to \mathbb{N}.$$
The arity of a symbol $x$ is the number of arguments referenced 
by $x$.
 %We assume that for each $n\in \Bbb{N}$ there are infinitely many elements of $\f$ and $\p$ of arity $n$, so we never run out of symbols of any required arity.
 %The $\arity$ function corresponds to the notion of `similarity type' in modal logic.  
 Note that in the case when the arity of a function symbol $g$ is zero,  
 $g$ may be thought of as a constant. 

 \begin{definition}[schemas]\label{struc.sch.def} \rm
We define the set  of all \emph{schemas} recursively as follows. 
$l: \ski$ is a schema. 
An   assignment $l: y\as f(\vx);$ where $y\in\var$, $f\in \f$, $l\in \labels$  and $\vx$ is a vector of $\arity(f)$ variables,  is a schema. Similarly an equality \ass\ $l: y \as x;$ for $y,x\in \var$ is a schema. 
From these all schemas 
may be `built up' from the 
following constructs on schemas. 

\begin{description}
\item[sequences] 
$S'=U_1 U_2\ldots U_r $ is a schema  provided that 
 each $U_i$ for $i\in \{1,\ldots, r\}$ is a  schema.  

\item[{\it if} schemas] $S''= l: \si  {p}(\vx) \then T_1 \els  T_2$ is a schema 
 whenever $p\in \p$, $l\in \labels$, $\vx$ is a vector of $\arity(p)$ variables, 
and $T_1,T_2$ are schemas.

% We call the schemas $T_1$ and $T_2$  the \emph{true} and \emph{false} parts  of  $p\labl$. and write $\part[S'']{\tru}({p})$ and $\part[S'']{\fal}(\ls{p})$ for $T_1$ and $T_2$ \resp.

\item[non-deterministic branches] $S''= l: T_1 \sqcup T_2 \ldots \sqcup T_m$ is a schema 
 whenever  $l\in \labels$ and $T_1, \ldots T_m$ are schemas.

\item[{\it while} schemas] $S'''= l: \whi {q}(\vy) T$ is a schema  
whenever $q\in \p$, $l\in \labels$, $\vy$ is a vector of $\arity(q)$ variables,
and $T$ is a schema. 

\begin{comment}
We call $T$ the \emph{body} of the \emph{while} predicate  $q$ in $S'''$. If $x$ is a labelled symbol in $T$, and there is no labelled while predicate $\ls[m]{p}$ in $T$ which also contains $x$ in its body, then we say that $\ls{q}$ lies \emph{immediately} above $x$. 
%, and write $\body[S'''](\ls{q})$ for $T$. 
\end{comment}

\item[non-deterministic loops] $S'''= l: \lop T $ is a schema if $l\in \labels$ and $T$ is a schema.

\item[{\it concurrent} schemas] $S''''= l: T_1 \conc T_2 \ldots \conc T_m$ is a schema, where $T_1, \ldots T_m$ are schemas.

\end{description}
\end{definition}

We only consider schemas without repeated labels; for example, in the case of the `while' schema $l: \whi {q}(\vy) T$, we assume that the label $l$ does not occur in the recursive definition of $T$.

The semantics of schemas are defined by their flowcharts, which are finite directed graphs. A directed graph $G$ is a pair $(V,E)$ with $E \subb V \times V$. We define $V= \vrtces(G)$, the set of \it{vertices} of $G$.

\begin{definition}  \label{graph.schema.defn} \rm

Given a schema $S$, we define a finite directed graph $\gr(S)$ 
%with vertex set $\labs(S) \cup \{\strt, \nd\}$
 with an edge labelling function $\edgtyp_S$ that associates to each edge  of $\gr(S)$ either $\emptwrd$, a triple $(p,\vx,X)$ for a predicate $p$, a vector $\vx$ of variables and $X\in \tf$,  or an \ass,\ 
as follows. Unless otherwise stated below, $\edgtyp_S$  maps to $\emptwrd$.

\begin{enumerate}

\item
 
If $S$ is $l: \ski$ or $l: y \as f(\vx);$ or $l: y \as x;$ then $\gr(S)$ has vertex set $\{\strt,l,\nd\}$ and   edges $(\strt,l)$ and $(l,\nd)$. Here $\edgtyp_S(l,\nd)= \varepsilon$, $ y \as f(\vx);$ or $ y \as x;$, respectively. 
%Also, if $S$ is $l: y \as f(\vx);$ or $l: y \as x;$, then $\edgtyp_S(l,nd)= \ul{y \as f(\vx)}$ or $\ul(y \as x)$, respectively. 

\item 

If $S= S_1 S_2$, then $\gr(S)$ has vertex set $$\vrtces(\gr(S_1)) \cup \vrtces(\gr(S_2))$$ and
contains every  edge occurring in either $S_1$ or $S_2$, with the function $\edgtyp_S$  returning the same value as in $S_1$ or $S_2$ respectively, except that $\gr(S)$ does not have any edge $(l,\nd)$ for a vertex $l$ in $S_1$ or $(\strt,l)$ for  a vertex $l$ in $S_2$. Instead, it has 
 an edge $(l_1,l_2)$ for each pair of edges $(l_1,\nd)$ and $(\strt,l_2)$ in $\gr(S_1)$ and $\gr(S_2)$ respectively, with the function $\edgtyp_S(l_1,l_2)=\edgtyp_{S_1}(l_1,\nd)$.

\item
 
If $S= l: S_1 \sqcup S_2\ldots \sqcup S_m$, then 
$\gr(S)$ has vertex set $$\vrtces(\gr(S_1)) \cup \ldots \cup \vrtces(\gr(S_m)) \cup \{l\}$$ and contains all edges $(l',l'')$ lying in any $\gr(S_k)$ \st\ $l'\not= \strt$,  with the function $\edgtyp_S$ returning the same value as $\edgtyp_{S_k}$in the appropriate $\gr(S_k)$, and also contains an edge $(l,l'')$ for each edge $(\strt,l'')$ in any $\gr(S_k)$. Additionally, $\gr(S)$ contains the edge $(\strt,l)$.

\item[$(3')$] 
If $S= l: \si p(\vx) \then S_1 \els S_2$, then 
$\gr(S)$ is identical to $\gr(l: S_1 \sqcup S_2)$ except that the edges $(l,l'')$ for each edge $(\strt,l'')$ in either $\gr(S_1)$ or $\gr(S_2)$ are mapped by $\edgtyp_S$ to $(p,\vx,\tru)$ or $(p,\vx,\fal)$ respectively.

\item
 
If $S= l: \whi q(\vy)  T $, then  $\gr(S)$ has vertex set $\vrtces(T) \cup \{ l\}$ and contains all edges $(l',l'')$ lying in $\gr(T)$  \st\ $l'\not= \strt$ and $l''\not= \nd$, with the functions $\edgtyp_S$  returning the same value as $\edgtyp_T$, and also contains an edge $(l,l'')$ for each edge $(\strt,l'')$ in $\gr(T)$, with $\edgtyp_S(l,l'') = (q,\vy,\tru)$,  and an edge $(l'',l)$ for each edge $(l'', \nd)$ in $\gr(T)$, with  $\edgtyp_S(l'',l) = \edgtyp_T(l'', \nd)$. Additionally, $\gr(S)$ contains the edges $(\strt,l)$ and $(l,\nd)$, with  $\edgtyp_S(l,\nd)=(q,\vy,\fal)$.

\item[$(4')$] 
 If $S= l: \lop T $, then $\gr(S)$ is identical to $\gr(\whi  q(\vy)  T)$, except that edges with $l$ as initial vertex map to $\emptwrd$ under $\edgtyp_S$.

\item 

If $S= l: \,S_1 \conc S_2 \conc \ldots \conc S_m$, then $\gr(S)$ has vertex set
 $$\times_{i=1}^m (\vrtces(\gr(S_i)) \cup \{\strt,l,\nd\},$$ and given any $r\le m$ and vertices $l_i \in \vrtces(\gr(S_i))$ for all $
i \not= r$ and any  edge $(l',l'')$ in $\gr(S_r)$, the graph $\gr(S)$ has an edge $$((l_1,\ldots,l_{r-1}, l', l_{r+1}, \ldots, l_m), (l_1,\ldots,l_{r-1}, l'', l_{r+1}, \ldots, l_m))$$ whose image under $\edgtyp_S$  is equal to
$\edgtyp_{S_r}(l',l'')$. 
 Additionally, \\ 
$\gr(S)$ contains the edges $(\strt,l)$, $(l,(\strt,\ldots,\strt))$ and \\ $((\nd,\ldots,\nd),\nd)$.

\end{enumerate}

\end{definition}

 \subsection{Semantics of schemas} \label{struc.sch.sem.subsec} \rm

The symbols upon which schemas are built are given meaning by 
defining the notions of a state and of an interpretation. It 
will be assumed that variables take values in the set of terms built from the sets of variables and function symbols. This set, which we denote by $\term(\f,\var)$, is usually called the Herbrand domain.
It is formally defined as follows: 
\begin{itemize}
\item each variable is a term, 
\item
if $ f\in {\f}$ is of arity $
n$ and $ t_1,\ldots ,t_n$ are terms then $ f(t_1,\ldots ,t_n)$ is a term. 
\end{itemize}
 The function symbols represent the `natural' functions \wrt\ the set of terms; that is, each function symbol $f$ defines the function $(t_1,\ldots,t_n) \mapsto  f(t_1,\ldots,t_n)$ for all $n$-tuples of terms $(t_1,\ldots,t_n)$. 
  A \emph{state} is  a function from $\var$ into the set of terms. An interpretation $i$ defines, for each
predicate symbol $p\in\p$ of arity $m$, a function 
$ p^i:\term(\f,\var)^m \ra \{\tru ,\, \fal \}$. 
We define 
the \emph{natural state} $e:\var \ra \term(\f,\var)$ 
 by $e(v)=v$ for all $ v\in \var.$

\begin{definition}[state associated with a path through $\gr(S)$ for schema $S$]\rm

Given a state $d$, a schema $S$ and a path $\nu$ through $\gr(S)$ whose first element is $\strt$, we define the state $\dd{\nu}{}$ recursively as follows.

\begin{itemize}

\item

$\dd{\strt}{}(v)= d(v)$ for all variables $v$.

\item 

If $\nu= \mu l l'$ for vertices $\l,l'$ in $\gr(S)$ and $\edgtyp(l,l')$ is not an \ass,\  then $\dd{\nu}{}=\dd{\mu l}{}$.

\item 

If $\nu= \mu l l'$ for $\l,l'\in \labs(S)$ and $S$ and $\edgtyp(l,l')= y\as f(x_1, \ldots, x_n);$, then $\dd{\nu}{}(z)=\dd{\mu l}{}(z)$ for all variables $z \not= y$, and $$\dd{\nu}{}(y)= f(\dd{\mu l}{}(x_1), \ldots ,\dd{\mu l}{}(x_n)),$$ and the case of equality \asss\ is treated analogously.

\end{itemize}

\end{definition}

%\begin{comment}
\begin{definition}[executable paths and free schemas]\rm  \label{exe.path.free.sch.defn}

Given a schema $S$ and  an \ter\ $i$ and  a path $\nu$ through $\gr(S)$ whose first element is $\strt$, we say
that $\nu$ is compatible with $i$ if 
 given any prefix $\mu l l'$ of $\nu$ \st\ $\edgtyp_S(l,l')=(p,x_1, \ldots, x_n,X)$, 
$p^i(\ee{\mu l}{}(x_1), \ldots ,\dd{\mu l}{}(x_n))= X$ holds. A path whose first element is $\strt$ is said to be executable if \txs\ an \ter\ with which it is compatible. A schema is said to be free if every path whose first element is $\strt$ is executable.
\end{definition}
%\end{comment}

Since a schema $S$ may contain the non-deterministic $\lop, \sqcup$ and $\conc$ constructions, an initial state $d$ and an \ter\ $i$ need not define a unique executable path in $\gr(S)$ from $\strt$ to $\nd$. In the event that only one executable path exists, we denote it by $\pathacm{S}{i,d}$, and write $\dd{S}{i}$ to  denote the state   $\dd{\pathacm{S}{i,d}}{i}$. If $S$ is merely a sequence of \asss, so that the \ter\ $i$ is irrelevant, then we simply write $\dd{S}{}$.

\subsection{The data dependence problems}

We now formalise the two data dependence conditions with which we are concerned in this paper.

\begin{definition} \rm 

Let $S$ be a schema and let $v\in \var$, let $l \in \vrtces(\gr(S))$ and let $f\in \f \cup \var$. The predicate $\sed(f,v,l)$ is defined to hold if there is an executable path $\mu$ through $\gr(S)$ which starts at $\strt$ and ends at $l$  \st\ the term  $\ee{\mu}{}(v)$ contains  $f$; and the predicate $\sad(f,v,l)$ is defined to hold if for every  executable path $\mu$ through $\gr(S)$ that starts at $\strt$ and ends at $l$, the term  $\ee{\mu}{}(v)$ contains  $f$.
 \end{definition}

%Thus both problems are stated in terms of the natural state $e$ as initial state. This id justified by the fact that $e$ is, in a sense, the most 

\section{Complexity Results for schemas having only equality \asss}

In this section, we prove that even if we restrict ourselves to the class of schemas without concurrency constructs and having only equality \asss,\ both the existential and universal data dependence problems are PSPACE-hard, and become NP-hard and co-NP-hard respectively if schemas are also required to be loop-free. We also show that if we keep the restriction to equality \asss\ but allow concurrency constructs, and add the further assumption of a constant bound on the arity of any predicate symbol, both problems lie in PSPACE.

\subsection{Notational conventions} 
\begin{itemize}
\item 
In the proof of Theorems \ref{hardness.np.sed.thm} and \ref{hardness.pspace.sed.thm}, we will define schemas without indicating labels, and indicate paths simply by using sequences of predicates and $\nd$. These schemas do not have the concurrency $\conc$ symbol and hence all vertices in the appropriate graph $\gr(S)$ lie in $\labs(S) \cup \{\strt, \nd\}$. In the cases where this convention is used, paths in the sense of Definition \ref{graph.schema.defn} are defined unambiguously.
\item
We will need to refer to finite sets of non-negative integers `without gaps'. Thus we define the set 
$$ [m,n] = \lbrace m, m+1, \ldots, n \rbrace$$ for any $m \le n $. 
\item
In order to save space, we will sometimes abbreviate schemas consisting of sequences of equality \asss\  by using the quantifier $\forall$. For example, in Fig. \ref{automaton.lem.fig}, the line $\forall k\in [0,m_j] \; t_k \as s_{j,k}$  is intended as a shorthand for the sequence $$t_0 \as s_{j,0};t_1 \as s_{j,1}; \ldots ;t_{m_j} \as s_{j,m_j};$$ 
The lines $\forall j\in  [1,m]  \;   \forall k \in [0,m_j]\; s_{j,k}\as u_{bad};$ and $ \forall s\in  \bigcup_{j \in [1,m]} \!  F_j\;\; s\as u_{good};$ in Fig. \ref{pspace.hard.fig} have  analogous meanings. 
We only use this notation in cases where the order of the \asss\ is immaterial, since no variable occurs on both the left side of one \ass\ in the sequence and the right side of another, and so the \asss\ commute. 
\item
In Lemma \ref{automaton.lem} and Theorem \ref{hardness.pspace.sed.thm}, we will define finite state automata for which the word `state' has its usual meaning; however we will also define schemas having variables which are the states of the automata, and thus the word state has the distinct meaning of a function from variables (automata-theoretic states) to elements of the domain (variables, in the case of schemas having only equality \asss).\ This should not cause confusion. 
\end{itemize}

\subsection{NP-hardness of data dependence problems for loop-free schemas without concurrency constructions}

Our main NP-hardness result follows.

\begin{theorem} \label{hardness.np.sed.thm}
For a schema $S$,  $v\in \var$ and  $f\in  \var$,
the problem of deciding  $ \sed(f,v,\nd)$ is NP-hard and that of deciding  $ \sad(f,v,\nd)$ is co-NP-hard,  even when (in the case of both problems) $S$ is restricted to membership of the class of schemas satisfying the following conditions. 
\begin{itemize}
\item
$S$ has no concurrency or non-deterministic branching constructions and has only equality \asss.
\item
 $S$ contains no loops.
 \item
 Each predicate in $S$ has zero arity. 
\end{itemize}
\end{theorem}

\proof 
We consider $\sed$ first, and then indicate the proof for $\sad$. 
To show NP-hardness of deciding whether $ \sed(f,v,l)$ holds, we use a polynomial-time reduction from 3SAT, which is known to be an NP-hard problem \cite{cook:3sat}. An instance of 3SAT comprises a set $X =\{x_1,\ldots, x_n\}$ and  a propositional formula $\rho=\bigwedge_{k=1}^{m}y_{k,1} \vee y_{k,2} \vee y_{k,3}$, where each $y_{i,j}$ is either $x_k$ or $\neg x_k$ for some $k\le n$. The problem is satisfied if \txs\ a valuation $\delta: X \cup \neg X \to \tf$ \st\ for each $x\in X$, $\{\delta(x), \delta(\neg x)\}= \tf$, under which $\rho $ evaluates to $\tru$. 
 Given this instance of 3SAT we  will construct a schema $S$ that satisfies the conditions given in the statement of the Theorem and  contains  variables $u_{bad},u_0, \ldots ,u_n$  \st\ $\sed(u_0,u_n,\nd)$ holds \ifa\ $\rho $ is satisfiable. 
The schema $S$ is 
$$  \forall j \in  [1,n]  \; u_j \as u_{bad}; \,\, D_1\ldots D_m,$$ where $D_l$ is as defined in Figure \ref{np.hard.fig}. Clearly $S$ can be constructed in polynomial time from the given instance of 3SAT, as required.

\begin{figure}[t]

\begin{center}$\begin{array}{llllll}      
        &   D_l \equiv C_{l,y_{l,1}}   C_{l,y_{l,2}} C_{l,y_{l,3}}  \\
\\
\text{where}& C_{l,y} \equiv \begin{cases} \si q_j() \then  u_l \as u_{l-1}; & \text{if } y = x_j\\
                                            \si q_j() \then \ski \els  u_l \as u_{l-1}; &  \text{if } y = \neg x_j
                                     \end{cases}
    \end{array}$
\end{center}
\caption{The  definition of the schema $D_l$ used in the proof of  Theorem \ref{hardness.np.sed.thm}.} \label{np.hard.fig} \vspace{10pt}
\end{figure}

% \rule{\linewidth}{.5pt}

Assume first that \txs\ a valuation $\delta: X \to \tf$ under which $\rho $ evaluates to $\tru$. Define the \ter\ $i$ to map $q_j()$ to $ \delta(x_j)$ for each $q_j$. Then the path $\pathacm{S}{i,e}$ clearly passes through at least one \ass\ $u_l \as u_{l-1}; $ within each $D_l$ in $S$, proving $\sed(u_0,u_n,\nd)$ holds. Conversely, if $\sed(u_0,u_n,\nd)$ holds, then there is an \ter\ $i$ \st\ the path $\pathacm{S}{i,e}$ passes through the sequence of \asss\ $u_1 \as u_{0}; $, \ldots, $u_n \as u_{n-1}; $ in turn, and hence passes through $u_l \as u_{l-1}; $ at least once within each $D_l$. Define the valuation $\delta$ as follows; $\delta(x_j)= \tru$ \ifa\ $i$ maps $q_j()$ to $\tru$. Clearly $\rho $ evaluates to $\tru$. Thus we have  proved  the Theorem for $\sed$. 

To prove co-NP-hardness of deciding the $\sad$ relation under the restricted conditions given, observe that the final value of the variable $u_n$ always lies in $\{u_0, u_{bad}\}$ and so $\sed(u_0,u_n,\nd) \iff  \neg \sad(u_{bad},u_n,\nd)$ holds. Thus deciding $\sad(f,v,\nd)$ is co-NP-hard.
\endproof

\subsection{PSPACE-hardness result for data dependence problems for schemas without concurrency constructions}

The main theorem of this subsection, Theorem \ref{hardness.pspace.sed.thm}, uses a polynomial-time reduction from the following automata-theoretic problem.

\begin{definition} \rm \label{inter.auto.kozen.defn}
Consider a set of  deterministic finite state automata $A_1,\ldots,A_m$ for some $m\ge 0$, all using an alphabet $\Sigma$.
The finite state automata intersection problem is that of deciding whether there exists a word in $\Sigma^*$ that is accepted by every automaton $A_j$.
\end{definition}

\begin{theorem}[\cite{kozen.finite.auto.intersect}] \label{kozen.inter.auto.thm}
The finite state automata intersection problem is PSPACE-complete.  \endproof
\end{theorem}

Given a deterministic finite state automaton $A$ and a member $\sigma$ of its alphabet, we wish to construct a schema consisting only of a sequence of \asss\  whose variables
are the states of $A$ and 
 such that for any transition $s \overset{\sigma}{\rightsquigarrow}  s'  $  in $A$,  $ \sad(s',s,\nd)$ holds. The schema 
 $$
\begin{array}{llllll}      
                           
                               &&\forall k\in  [0,a] \; t_k \as s_{k};\\
                                      &&        \forall k\in [0,a] \;  s_{k} \as t_{\chi(k)};
                                         \end{array}$$
satisfies this requirement if $A$ has state set $\lbrace s_0, \ldots, s_a \rbrace$ and its $ \sigma$-transitions all have the form $s_k \overset{\sigma}{\rightsquigarrow}  s_{\chi(k)}  $ for a function $\chi:  [0,a] \to                          [0,a]  $, with new variables $t_k$ disjoint from the variables $s_l$. It may be worth mentioning that the simpler schema 
$$\begin{array}{ll} & s_0 \as s_{\chi(0)};\\
                    & \vdots \\
                    & s_a \as s_{\chi(a)}; 
                    \end{array} $$ 
does \emph{not} satisfy the required data dependence condition because the \asss\ may `interfere' with one another; for example, if $A$ has only two states $s_0,s_1$ and has transitions 
$s_1 \overset{\sigma}{\rightsquigarrow}  s_0  $ and $s_0 \overset{\sigma}{\rightsquigarrow}  s_1  $, then if $S$ is the schema
                   $$\begin{array}{ll} & s_0 \as s_1;\\
                     & s_1 \as s_0; 
                    \end{array} $$ 
then $ \sad(s_1,s_1,\nd)$ rather than the required  $ \sad(s_0,s_1,\nd)$ holds. Thus it is necessary to introduce the `copying' variables $t_k$.                   
                    
The motivation for constructing a schema in this way from a given finite state automaton is shown by Lemma \ref{automaton.lem}.

\begin{lemma} \label{automaton.lem}
 
Consider a set of $m$ deterministic finite state automata $A_1,\ldots,A_m$ for some $m\ge 0$, all using an alphabet $\Sigma=\{\alpha_1,\ldots, \alpha_n\}$, with each automaton  $A_j$ having state set  $S_j=\{s_{j,0},\ldots, s_{j,m_j}\}$ and total transition function  $\eta_j:\Sigma \times S_j \to S_j$.
For each automaton  $A_j$ and each letter $\alpha_l \in \Sigma$, let $U_{j,l}$ be the predicate-free schema in Fig. \ref{automaton.lem.fig} and define $V_l = U_{1,l} \ldots U_{m,l}$. 
Let $l_1,l_2,\ldots, l_r \in   [1,n]  $ and define $\gamma = \alpha_{l_r}\alpha_{l_{r-1}}\ldots \alpha_{l_1}\in \Sigma^*$. 

\begin{enumerate}
\item
For every $j \in  [1,m]  $ and any $s \in S_j$, $\sad[V_{l_1} \ldots V_{l_r}](\eta_j(\gamma,s),s,\nd)$ holds.
\item 
Assume each automaton  $A_j$ has initial state $s_{j,0}$ and   final state set $F_j \subb S_j$. 
Let $e_{final}$ be the state (in the program sense)
$$\begin{cases} s_{j,k} \mapsto u_{bad} & s_{j,k} \in S_j -F_j\\
                s_{j,k} \mapsto u_{good} & s_{j,k} \in   F_j
\end{cases}$$

for new variables $u_{bad}$, $u_{good}$.
Then  $$\mean{V_{l_1} \ldots V_{l_r}}{}{e_{final}}(s_{j,0})= u_{good}$$ for all $j$ \ifa\ the word $\gamma$ is accepted by every automaton $A_j$.
\end{enumerate}
\end{lemma}

\begin{figure}[t]

\begin{center}
$ 
\begin{array}{llllll}      
                           
                               &&\forall k\in  [0,m_j] \; t_k \as s_{j,k};\\
                                      &&        \forall k\in  [0,m_j] \;  s_{j,k} \as t_{\chi_j(l,k)};
                                         \end{array}$
                                        
\end{center}
\caption{The schema $U_{j,l}$ of Lemma \ref{automaton.lem}. Here the $t_k$ are new variables used solely for copying and the function $\chi_j$ is defined by the state transition function $\eta_j$ of the automaton $A_j$ as follows; for any letter $\alpha_l$ and state $s_{j,k}$, $\eta_j(\alpha_l, s_{j,k}) =s_{j,\chi_j(l,k)}$. Observe that the value defined by a variable $s_{j,k}$ after execution of $U_{j,l}$ is the same as that defined by the variable $\eta_j(\alpha_l, s_{j,k})$ before execution, since $ \sad[U_{j,l}](\eta_j(\alpha_l, s_{j,k}),s_{j,k},\nd)$ holds.} \label{automaton.lem.fig} \vspace{10pt}
\end{figure}

\proof
(1)  can be straightforwardly proved by induction on $r$. (2) follows immediately from (1) using the fact that for any $j$, $A_j$ accepts $\gamma$ \ifa\ $  \eta_j(\gamma,s_{j,0}) \in  F_j$ holds.\endproof

We now give 
the main PSPACE-hardness theorem of the paper, Theorem \ref{hardness.pspace.sed.thm}. The proof of this Theorem  will construct a schema in which solving an existential data dependence 
problem corresponds to solving a given instance of the finite state automata
intersection problem. Parts of the schema constructed will `simulate' state
transitions of the automata.

\begin{theorem} \label{hardness.pspace.sed.thm}
For a schema $S$,  $v\in \var$ and  $f\in  \var$,
the problems of deciding  whether  $ \sed(f,v,\nd)$ and $ \sad(f,v,\nd)$ hold are both PSPACE-hard, even when $S$ is restricted to membership of the class of schemas satisfying the following  conditions.
\begin{itemize}
\item
$S$ has no concurrency or non-deterministic branching constructions and has only equality \asss,\
\item
 No predicate occurs more than once in $S$.
 \item
 $S$ contains two while predicates, one of which lies in the body of the other.
\end{itemize}
\end{theorem}

\proof 
We consider $\sed$ first, and then indicate the proof for $\sad$. 
We  prove the Theorem using a reduction from the intersection problem for finite state automata, given in Definition \ref{inter.auto.kozen.defn}, which is PSPACE-complete by Theorem \ref{kozen.inter.auto.thm}. Thus we assume an instance of this problem comprising a set of $m$ deterministic finite state automata $A_1,\ldots,A_m$ for some $m\ge 0$, all using an alphabet $\Sigma=\{\alpha_1,\ldots, \alpha_n\}$, with each $A_j$ having state set  $S_i=\{s_{j,0},\ldots, s_{j,m_j}\}$,  total transition function  $\eta_j:\Sigma \times S_j \to S_j$, initial state $s_{j,0}$ and final state set $F_j \subb S_j$, as in the statement of Lemma \ref{automaton.lem}. The problem is satisfied if there is a word in $\Sigma^*$ which is accepted by every automaton $A_j$. 

Given these automata, consider the schema $S$ given in Fig. \ref{pspace.hard.fig}. Clearly $S$ satisfies  the conditions listed in the statement of the Theorem 
  and $S$ can be constructed in polynomial time from the set of automata $A_j$ as input. 
 We now show that 
  $\sed(u_{good},a_m,\nd)$ holds \ifa\ the intersection of  the acceptance sets of all the automata $A_j$ is non-empty,   thus proving the Theorem.

 \begin{figure}[h]

\begin{center}$ \begin{array}{llllll}  \forall j \in  [1,m] \; a_j \as u_{bad}; \\
                                         \forall j  \in  [1,m]   \; b_j \as u_{bad}; \\
     
                                         \whi Q_1(a_m) &\{ \\
                                       &      \forall j  \in [1,m] \; a_j\as u_{bad};\\
                                        & \forall j  \in  [1,m-1] \; b_j\as u_{bad};\\
                                        & c\as u_{bad};\\
                                                    & \forall j\in [1,m] \;
                                                    \forall k \in   [0,m_j]  \;\; s_{j,k}\as u_{bad};\\
                                         & \si Q_2(b_m)  &  \!\!\!\!  \!\!  \then & c \as u_{good};\\
                                                \\
                                         &        &       \!\!\!\!  \!\!    \els      & \{  \\
							    &&&  \forall s\in  \bigcup_{j \in [1,m]} \!  F_j\;\; s\as u_{good};\\
							    &&&\whi Q_3(s_{1,0},\ldots,s_{m,0})\;\;\;T_n\\
                                                  &        &        & \}  \\
                                                   \\
						&  \si p_1(s_{1,0}) & \!\!\!\! \!\! \then & a_1 \as b_m;\\
                                                  &&\!\!\!\!\!\!   \els & b_1 \as c; \\
                                                   \\
                                             & \si p_2(s_{2,0}) & \!\!\!\! \!\! \then &  a_2 \as a_1; \\ 
                                                & & \!\!\!\!  \!\! \els &   b_2 \as b_1; \\
                                              & \vdots \\
                                              &   \vdots \\
                                            & \si p_m(s_{m,0}) & \!\!\!\! \!\! \then &  a_m \as a_{m-1}; \\ 
                                                & &  \!\!\!\! \!\! \els &   b_m \as b_{m-1}; \\
						&\}\\
                                         \end{array}$
\end{center}
\caption{The schema $S$ used in the proof  of Theorem \ref{hardness.pspace.sed.thm}. The schema $T_n$ is defined in Fig. \ref{Tl.auto.fig}.} \label{pspace.hard.fig} \vspace{10pt}
\end{figure}

\begin{figure}[h]

\begin{center}$\begin{array}{llllll}

                   &T_l \equiv     \;\;\;\; \;\;\;  &  \si q_l(\vs) & \then &  V_l \\
                   &                   &               & \els  & T_{l-1} \\
\\
                   & T_1 \equiv V_1\\
\\
 &\text{where}&  V_l \equiv U_{1,l} \ldots U_{m,l}
                                         \end{array}$
\end{center}
\caption{The recursive definition of the schema $T_l$. Here $\vs$ is a vector whose entries are all the variables $s_{j,k}$, in any fixed order, and $U_{j,l}$ is the schema in Fig. \ref{automaton.lem.fig}. Observe that an execution of $T_n$ entails an execution of one schema $V_l$, for some $l \in   [1,n]   $.  \label{Tl.auto.fig}   }  \vspace{10pt}
\end{figure}

\begin{itemize}

\item ($\La$). 
Assume first that there is a word $\gamma = \alpha_{d_z}\alpha_{d_{z-1}}\ldots \alpha_{d_1}$ that is accepted by every automaton $A_j$, for minimal $z$. We will prove that $\sed(u_{good},a_m,\nd)$ holds. 
Define the \ter\ $i$ on the predicates $Q_1,Q_2,Q_3$ and each $p_j$ as follows.  
$$\begin{cases} Q_1(u_{bad}) \mapsto \tru,\;\; Q_1(u_{good}) \mapsto \fal\\
                Q_2(u_{bad}) \mapsto \tru,\;\; Q_2(u_{good}) \mapsto \fal \\
                Q_3(v_1,\ldots, v_m) \mapsto \fal \text{ iff every } v_j= u_{good}\\
                p_j(u_{bad}) \mapsto \fal, \;\; p_j(u_{good}) \mapsto \tru\\
\end{cases}$$ 

We now indicate how $i$ is defined on the predicates $q_l$. Define the path
$$ \begin{array}{ll} \mu  &= Q_3 q_n q_{n-1}\ldots q_{d_1}Q_3 q_n q_{n-1}\ldots q_{d_2}Q_3\ldots Q_3 q_n q_{n-1}\ldots q_{d_z}Q_3 \\ &\in\pathset(\whi(Q_3(s_{1,0},\ldots,s_{m,0})) \, T_n).\end{array}$$

We wish $\pathacm{S}{i,e}$ to follow the path $\mu p_1$ whenever it encounters 
\\ $\whi(Q_3(s_{1,0},\ldots,s_{m,0})) \, T_n)$, in effect executing the schema $V_{d_1} \ldots V_{d_z}$. We now show that this is possible. First observe that by Part (2) of Lemma \ref{automaton.lem} applied to the suffices of $\gamma$, every variable   $s_{j,0}$ defines the value  $u_{good}$ at the last occurrence of $Q_3$ along $\mu$, but this does not hold at any earlier occurrence of $Q_3$, since this would imply that a strict suffix of $\gamma$ was accepted by every automaton $A_j$, contradicting the minimality of $z$. Thus the definition of $i$ on $Q_3$ given above ensures that $\pathacm{S}{i,e}$  follows the path $\mu p_1$ where required, provided that $i$ can defined appropriately on each predicate $q_l$.

Suppose that this is impossible; that is, that there is a repeated $q_l$-predicate term along $\mu$ for some $q_l$, which $i$ would have to map to both $\tru$ and $\fal$. Thus we can write 
$\mu= \mu' q_l \mu'' q_l \mu'''$ \st\ every variable $s_{j,k}$ defines the same value at the two occurrences of $q_l$.
Assume that $Q_3$ occurs $z'$ times in $\mu'$ and $z''$ times in $\mu''$; clearly $z'' \ge 1$. 
Since no variable apart from the variables $s_{j,k}$ occurs in the while schema guarded by $Q_3$, every variable $s_{j,0}$ defines the same value after the path $\mu' q_l \mu''$ as after $\mu$, namely $u_{good}$. Thus  by Part (2) of Lemma \ref{automaton.lem}, the word 
$\alpha_{d_z}\alpha_{d_{z-1}} \ldots \alpha_{d_{z'+z''}}\ldots \alpha_{d_{z'-1}}\ldots \alpha_{d_1}$ is accepted by every automaton $A_j$, contradicting the minimality of $z$.

Thus we have shown that 
the \ter\ $i$ can be defined so that $\pathacm{S}{i,e}$ always follows the path $\mu$  whenever $\whi(Q_3(s_{1,0},\ldots,s_{m,0})) \, T_n$ is reached, and furthermore, every variable $s_{j,0}$ defines the value  $u_{good}$ at the end of $\mu$, and so $p_1$ is the next symbol though which $\pathacm{S}{i,e}$ passes.

We now prove that $\ee{S}{i}(a_m) = u_{good}$ holds. The definition of $i$ on $Q_1$ ensures that $\pathacm{S}{i,e}$ passes at least once through the body of $Q_1$, and since $i$ maps $Q_2(b_m)$ to $\tru$ and each $p_j(u_{bad})$ to $\fal$, on the first passing of $\pathacm{S}{i,e}$ through the body of $Q_1$, the \ass\ $c\as u_{good};$ and  all \asss\ to every $b_j$ occur, and hence $b_m$ defines the value $u_{good}$ when $Q_1$ is reached for the second time along $\pathacm{S}{i,e}$. Since $i$ maps $Q_1(u_{bad})$ to $\tru$, the path $\pathacm{S}{i,e}$ then enters the body of $Q_1$ a second time, and since $i$ maps $Q_2(u_{good})$ to $\fal$, this time $\pathacm{S}{i,e}$ passes through $Q_3$. As proved above, $\pathacm{S}{i,e}$ terminates within $\whi(Q_3(s_{1,0},\ldots,s_{m,0})) \, T_n)$ and 
every $s_{j,0}$ defines $u_{good}$ when $\pathacm{S}{i,e}$ then reaches $p_1$, and so $\pathacm{S}{i,e}$ then passes through all the \asss\ $a_1 \as b_m;$ and $a_j \as a_{j-1};$, after which $a_m$ defines the value $u_{good}$. Since $i$ maps $Q_2(a_m)$ to $\fal$, 
$\sed(u_{good},a_m,\nd)$ holds, as required.

\item ($\Ra$). 
Conversely, suppose that $\sed(u_{good},a_m,\nd)$ holds. Thus $\ee{S}{i}(a_m) = u_{good}$ holds for some \ter\ $i$. The only sequence of \asss\ which could copy $u_{good}$ at the start of $S$ to $a_m$ at the end consists, in order, of the  \ass\   $c\as  u_{good};$ and those referencing every $b_j$ for $j <m$ followed by those  referencing  $b_m$ and every $a_j$ for $j <m$, and so $\pathacm{S}{i,e}$ must pass through all of these in turn. Furthermore, owing to the \asss\ setting $c$ and $b_1,\ldots, b_{m-1}$ to $ u_{bad}$, the \asss\ referencing $c$ and every $b_j$ for $j <m$ must occur in a single passing through the body of $Q_1$, during which every $s_{j,0}$ defines $u_{bad}$ when $p_j$ is reached. Thus $i$ must map every $p_j(u_{bad})$ to $\fal$. 
Similarly, owing to the \asss\ $a_j\as u_{bad};$,  the \asss\ referencing  every $a_j$ for $j <m$ must also occur in a single passing through the body of $Q_1$, and so the predicate term defined by each $p_j(s_{j,0})$ must map to $\tru$, and so every $s_{j,0}$ must define a value distinct from $u_{bad}$ simultaneously. The only possibility is $u_{good}$, and so at some point the path $\pathacm{S}{i,e}$, must reach $p_1$ with  each  $s_{j,0}$ defining $u_{good}$, and thus must have passed through $Q_3$ since the last occurrence of $Q_2$. Let $V_{d_1} \ldots V_{d_z}$ be the sequence of schemas $V_k$
occurring on $\pathacm{S}{i,e}$ since this occurrence; then by Part (2) of Lemma \ref{automaton.lem}, the word 
$\alpha_{d_z}\alpha_{d_{z-1}}  \alpha_{d_1}$ is accepted by every automaton $A_j$, as required.  
\end{itemize} 

To prove PSPACE-hardness of deciding the $\sad$ relation, observe that the final value of the variable $a_m$ always lies in $\{u_{good}, u_{bad}\}$ and so $\sed(u_{good},a_m,\nd) \iff  \neg \sad(u_{bad},a_m,\nd)$ holds. Thus deciding $\sad(f,v,\nd)$ is co-PSPACE-hard and hence PSPACE-hard. \endproof

%\addtolength{\arraycolsep}{10pt}

\subsection{Membership in PSPACE of data dependence problems for the class of schemas having a bound on the arity of all predicates and having only equality assignments, but without restrictions on concurrency constructs}

In order to prove that our problems lie in PSPACE, we need to show that the successors of a vertex in $\gr(S)$ can be enumerated in polynomial time. This motivates Theorem \ref{schema.encode.thm}.

\begin{theorem}\label{schema.encode.thm}

Let $S$ be a schema. 
\begin{enumerate}
\item 
The vertices of $\gr(S)$ can be encoded as words in the alphabet $\labs(S) \cup \{\strt,\nd\}$ in which no element of 
$\labs(S)$ occurs more than once and $\strt$ and $\nd$ each occur not more than $\vert \labs(S) \vert $ times.
\item
Given any $l' \in \vrtces(\gr(S))$, the set of all $l'' \in \vrtces(\gr(S))$ for which $(l',l'')$ is an edge in $\gr(S)$, and the corresponding values of $\edgtyp(l',l'')$,  can be computed in polynomial time.
\end{enumerate}
\end{theorem}

\proof
\begin{enumerate}
\item 
 We indicate the encoding by assuming that $S$ has the form $S= l: \,S_1 \conc S_2 \conc \ldots \conc S_m$; the encoding in the case of the other constructions given in Definition \ref{graph.schema.defn} is straightforward to infer. In the concurrent case, $\gr(S)$ has vertex set
 $\times_{i=1}^m (\vrtces(\gr(S_i)) \cup \{\strt,l,\nd\}$ and a vertex of  $\gr(S)$ can be encoded either by an element of $\{\strt,l,\nd\}$ (representing themselves) or by a word $w=w_1\ldots w_m$, where each $w_i$ represents an element
\\  $l_i \in \vrtces(\gr(S_i))$ and $w$ represents $(l_1,\ldots, l_m)$. The conditions given on the frequency of letters in $w$ follow easily from those for each $w_i$ and the fact we assume that no label occurs more than once in $S$. 
\item
This follows easily by induction on the structure of $S$, using the encoding given in Part (1) of this Theorem. \endproof
\end{enumerate}

Our other main theorem of this Section follows.

\begin{theorem} \label{pspace.all.eq.sed.thm}
Let $S$ be a schema and let $v\in \var$, let $l$ be a vertex of $\gr(S)$  and let $f\in  \var$. Assume that all \asss\ in $S$  are equality \asss.\ Assume that there is a constant upper bound on the arity of any predicate symbol occurring in $S$. Then the problems of deciding whether  $\sed[S](f,v,l)$ or $\sad[S](f,v,l)$ hold  both lie in PSPACE. 
\end{theorem}

\proof
 We first  prove decidability of $\sed[S](f,v,l)$  in PSPACE. We do this by
constructing the following algorithm, which lies in NPSPACE. 
 We non-deterministically guess a path beginning at $\strt$ through the schema $S$ that realises the copying of the initial value of the variable $f$ onto $v$ at the vertex $l$. At each point in the algorithm we store not just the vertex and the state (with the domain restricted to the set of variables referenced in $S$) reached, but also a finite, initially empty set of equations of the form $p(\vy)= X$ for predicate $p$ occurring in $S$, variable vector $\vy$ whose components are referenced in $S$ and $X\in \tf$. If $n$ is an upper bound on the total number of predicates and variables occurring in $S$ and $b$ is the assumed constant upper bound on the arity of any predicate in the class of schemas under consideration, then 
the number of equations of this form  is  bounded by $2n^{b+1}$ and thus the data stored at any point in the execution of the algorithm is polynomially bounded. 

 Whenever the algorithm crosses an edge $(l',l'')$ in $\gr(S)$ satisfying 
\\ $\edgtyp_S(l',l'')= (q,\vx,X)$, the equation $q(\vy)=X$ is added to the set, where the vector $\vy= \ee{\mu}{} \vx$, with $\mu$ being the path traced by the algorithm up to the vertex $l'$. No equation is added to the set when an edge for which   $\edgtyp_S$ returns $\epsilon$ or an \ass\ is crossed. Thus this equation set encodes the set of \ters\ which are compatible with the path followed, in the sense that an \ter\ $i$ is compatible with this path \ifa\ 
$p(\vy)= X$ is a consequence of $i$ for all equations $p(\vy)= X$ in the set.

 The algorithm terminates and returns {\it false} if the equation set acquires a pair of contradictory equations (that is, a pair $p(\vw)= \tru$, $p(\vw)= \fal$) at any point.
It terminates and returns {\it true} if  $l$ is reached with the state mapping $v$ to $f$ without two contradictory equations having occurred in the set.  By Theorem \ref{schema.encode.thm}, this algorithm lies in NPSPACE.  Since PSPACE $=$ NPSPACE holds, the problem of deciding $\sed[S](f,v,l)$ is thus in PSPACE.

To prove decidability of $\sad[S](f,v,l)$  in co-NPSPACE $=$ PSPACE instead, we modifiy the algorithm as follows; termination with output {\it true} occurs if $l$ is reached with the state \emph{not} mapping $v$ to $f$. 
\endproof

\section{Complexity Results for free schemas}

If we allow \asss\ with function symbols, and not just equalities, to occur in schemas, then deciding data dependence becomes harder, and the proof of membership in PSPACE for both problems in Theorem \ref{pspace.all.eq.sed.thm} does not appear to generalise. However, under  restriction to the class of free schemas, we prove in Theorem \ref{schema.free.withpreds.datdep.pspace.hard.thm} that deciding  existential data dependence is  PSPACE-complete, using  M\"uller-Olm's result   \cite{optimal.slic.parallel.progs.mueller-olm.seidl} for non-deterministic programs.
  Additionally, we prove in Theorem \ref{datdep.alg.poly.thm} that under the further condition that a constant bound is placed on the number of subschemas occurring in parallel,  this problem becomes polynomial-time decidable.  

Recall that a schema is \emph{free} if every path through its flowchart is executable. As an example, the schema 
$$  \whi {q}(z) \du z \as h(z);$$
(we have omitted labels from its definition) is free, whereas 
$  \whi {q}(z) \du z \as g();$
is not free, since there is no \ter\ and initial state such that the path so defined enters the body of $q$ exactly once.

\subsection{PSPACE-completeness of the existential data dependence problem for free schemas}

Theorem \ref{schema.free.withpreds.datdep.pspace.hard.thm} is the main result of this subsection.

\begin{lemma}\label{schema.nopreds.datdep.pspace.hard.lem}
Given any schema $S$ without predicates, a variable $v$ and $f \in \var \cup \f$, the problem of deciding whether  $ \sed[S](f,v,\nd)$ holds is PSPACE-hard. 
\end{lemma}

\proof
This is \cite[Theorem 2]{optimal.slic.parallel.progs.mueller-olm.seidl}. 
\endproof

\begin{lemma}\label{free.exist.schmea.lem}

Given any free schema $S$, a vertex $l$ in $\gr(S)$, a variable $v$ and $f \in \var \cup \f$, with $l$, $v$ and $f$ all occurring in $S$, there exists a free schema $S'$ which does not contain any $\lop$ or $\sqcup$ constructions, and such that $ \sed(f,v,l)$ holds \ifa\ $ \sed[S'](f,v,l)$ does. Furthermore, $S'$ can be constructed in polynomial time from $S$.
\end{lemma}

\proof
 Given $S$, we replace $\lop$ or $\sqcup$ constructions with while and nested if statements respectively, in the following way. Let $z$ be a variable not occurring in $S$ and not equal to $v$ or $f$, let $h$ be any function symbol and let $q$ be any  predicate symbol.  Suppose that $m: \lop T$ occurs in $S$; then we replace it by $m': z\as h(z);  m: \whi {q}(z) \du \{m'': z\as h(z); T\}$, for  new labels $m',m''$. Similarly, an occurrence of $m: T_n \sqcup \ldots \sqcup T_1$ in $S$ can be replaced by the schema $m: P_n$, where we recursively define $P_1 \equiv  z\as h(z);  T_1$ 
and $P_r \equiv \si q(z)   \then z\as h(z);  T_r \els  z\as h(z); P_{r-1}$ for $r >1$, where we have omitted labels in the definitions of each $P_r$. Let $S'$ be the schema obtained from $S$ after all the $\lop$ or $\sqcup$ constructions have been replaced. Since $z$ is never referenced in the original schema $S$,  the new \asss\ to $z$ cannot interfere with the existing data dependence relations in $S$, and the length of any term defined by $z$ along a path through $S'$ must successively increase at each \ass\ to $z$, hence the introduction of the new while and if statements cannot cause repeated predicate terms to occur. Thus $S'$ is free if $S$ is. There is a natural correspondence between paths in $S$ and in  $S'$, and thus $ \sed(f,v,l)$ holds \ifa\ $ \sed[S'](f,v,l)$ follows. 
Also, $S'$ can be constructed in polynomial time from $S$, proving the Lemma. \endproof

\begin{definition} \label{dat.dep.var.defn} \rm 
Given a schema $S$, $\,l,l' \in \vrtces(\gr(S))$ and variables $v,v'$, we define the relation $(l,v) \perc (l',v')$ to hold if either $\edgtyp(l,l')$ is an \ass\ to $v'$ that references $v$, or $v=v'$ and $\edgtyp(l,l')$ is not an \ass\ to $v'$.
\end{definition}

\begin{lemma}\label{dat.dep.var.trans.lem}

For any free schema $S$, a vertex $l$ in $\gr(S)$, a variable $v$ and $f \in  \f$,  $ \sed[S](f,v,l)$ holds \ifa\  \tx\ $m,n \in \vrtces(\gr(S))$ and a variable $w$ \st\ $\edgtyp(m,n)$ is an \ass\ to $w$ with  function symbol $f$ and $(n,w) \perc^* (l,v)$ holds.
\end{lemma}

\proof
This follows immediately from the definition of  $ \sed[S](f,v,l)$.
\endproof

\begin{theorem}\label{schema.free.withpreds.datdep.pspace.hard.thm}

Given any free schema $S$, a vertex $l$ in $\gr(S)$, a variable $v$ and $f \in \var \cup \f$, the problem of deciding whether  $ \sed[S](f,v,l)$ holds is  PSPACE-complete, and is  PSPACE-hard even if $l=\nd$ and $\,S$ does not contain any $\lop$ or $\sqcup$ symbols.  

\end{theorem}

\proof
 The PSPACE-hardness result follows immediately from Lemmas \ref{schema.nopreds.datdep.pspace.hard.lem} and \ref{free.exist.schmea.lem}. 

To show membership in PSPACE, we first assume that $f \in \f$, since if $f\in \var $ then we can replace  $S$ by the schema 
$S'  \equiv \, f\as g();\, S$ for a function symbol $g$ not occurring in $S$, for then 
$ \sed[S](f,v,l)\, \iff\,  \sed[{S'}](g,v,l)$ 
holds, and $S'$ can be constructed in polynomial time from the input. The result then follows from Lemma \ref{dat.dep.var.trans.lem}
 as follows. We non-deterministically guess an edge $(m,n)$ in $\gr(S)$ and a variable $w$ \st\ $\edgtyp(m,n)$ is an \ass\ to $w$ with  function symbol $f$ and then decide whether $(n,w) \perc^* (l,v)$ holds. This can be done by guessing a path from $(n,w)$ to $(l,v)$ in the digraph whose vertices are pairs $(l',v')$ for $l' \in \vrtces(\gr(S))$ and variables $v'$ occurring in $S$ and whose edges are given by the $\perc$ relation. At any point in the algorithm, only the current pair $(l',v')$ is stored, rather than the entire graph. By Theorem \ref{schema.encode.thm}, only polynomial space in the input is required for this, thus proving that the problem lies in NPSPACE $=$ PSPACE.
\endproof

\subsection{Polynomial-time complexity of the existential data dependence problem for the class of free schemas with a bound on the number of concurrency constructs}

We now consider the existential data dependence problem in which a constant upper bound is placed on the number of occurrences of $\conc$ in the schemas. Owing to the freeness assumption on the class of schemas under consideration,  $\sed$ can be defined by an iterative data flow analyis. 
Lemma \ref{poly.time.graph.lem} provides the crucial result in showing that in this case, the problem is polynomial-time bounded. This result relies on Lemma \ref{sch.flowchart.bound.lem}, which follows from the inductive definition of a schema flowchart in  Definition \ref{graph.schema.defn}.

\begin{lemma} \label{sch.flowchart.bound.lem}
Let $B$ be a non-negative integer and 
suppose that  there are non-decreasing functions $P_B: \Bbb{N} \to \Bbb{N}$ satisfying  the following conditions. 

\begin{description}
\item[$(0)$]
$P_{B+1}(n) \ge P_B(n)$ if $ n \ge 1$.
\item[$(1)$]  
$  P_B(n) \ge 3$ if  $ n \ge 1$. 

%\item[$(2)$]
%$ P_B(n_1 + n_2) \ge P_{C_1}(n_1) + P_{C_2}(n_2)$ if $C_1 + C_2 = B$ and $n_1, n_2 \ge 1$.

\item[$(2,3,3')$]
 
$   P_B(n_1 + \ldots + n_m ) \ge P_{C_1}(n_1)+\ldots + P_{C_m}(n_m) + 1$ if $m \ge 2$, $C_1 + \ldots + C_m = B$ and  $n_i \ge 1 \,\forall i$.
\item[$(4,4')$]
 
$ P_B(n+1) \ge P_B(n) + 1$ if $n \ge 1$.

\item[$(5)$]
 
$   P_B(n_1 + \ldots + n_m ) \ge P_{C_1}(n_1)\ldots P_{C_m}(n_m) + 3$  if $C_1 + \ldots + C_m = B - m+1$ and $B \ge m-1 \ge 1 $ and $ n_i \ge 1 \,\forall i $.
\end{description}
Then  for every schema $S$  encoded by a word of length $n$, in which  $\conc$ occurs  not more than $B$ times,  $\gr(S)$ has not more than $P_B(n)$ vertices.  
 \end{lemma}

\proof
This follows by induction on the structure of $S$. 
Each Condition in the statement of the Lemma apart from (0) is labelled with the number of the case in Definition \ref{graph.schema.defn} that requires it. As an example, consider Condition (5).  Assume that  $S= l: \,S_1 \conc S_2 \conc \ldots \conc S_m$; then $\gr(S)$ has vertex set \\
 $\times_{i=1}^m (\vrtces(\gr(S_i)) \cup \{\strt,l,\nd\}$. Assume that $\conc$ occurs not more than  $B'$ times in $S$ and exactly $C_i$ times in each $S_i$. Define $B = C_1 + \ldots + C_m + m -1$. Suppose each schema $S_i$ is encoded by a word of length $n_i$ and $S$ is encoded  by a word of length $n$, then  $$n \ge n_1 + \ldots + n_m $$ holds. By the inductive hypothesis, each $\gr(S_i)$ has not more than $P_{C_i}(n_i)$ vertices. Hence $\gr(S)$ has not more than $P_{C_1}(n_1)\ldots P_{C_m}(n_m) + 3 $ vertices, and hence by (5) and the monotonicity Condition (4), not more than $P_B(n)$ vertices. Thus since clearly $B' \ge B$ holds, it follows from (0) that $\gr(S)$ has not more than  $P_{B'}(n)$ vertices, proving the Lemma in this case.  Other cases are treated analogously. 
\endproof

\begin{lemma}
 \label{poly.time.graph.lem}
Given any integer $B\ge 0$, let $\chi_B$ be the set of all schemas in which  $\conc$ occurs  not more than $B$ times.   Then there exists an  algorithm that when given a schema $S$ in $\chi_B$ as input, constructs the graph $\gr(S)$ and is   polynomial-time bounded.
 \end{lemma}

\proof 
For each $B\ge 0$, it suffices to prove that the set containing \\ $\vert \vrtces(\gr(S))\vert$ for every schema $S$ in $\chi_B$
is polynomially bounded in terms of the number of letters needed to encode $S$.  The conclusion of the Lemma then follows from Part (2) of Lemma \ref{schema.encode.thm}. 
Consider the functions $P_{B}: n \mapsto max(3, n^{6(B+1)})$. 
 We will show that they satisfy Conditions (0--5) of Lemma \ref{sch.flowchart.bound.lem}, and hence that $P_{B}(n)$ is an upper bound for the number of vertices in $\gr(S)$ for any schema in $\chi_B$ encoded by a word of length $n$.  The existence of the polynomial  bound required will follow immediately. 

Clearly the functions $P_{B}$ satisfy Conditions $(0,1,4,4')$. We now prove that they satisfy Condition $(2,3,3')$ under the stated assumptions. 
Observe that 
$$P_B(n_1 + \ldots + n_m) = (n_1 + \ldots + n_m )^{6(B+1)} =  ((\sum_{i \le m} n_i)^2)^{3(B+1)}$$
$$ \ge  (\sum_{i \le m} n_i^2 +2n_1 n_2)^{3(B+1)} \ge (\sum_{i \le m} n_i^2 +1)^{3(B+1)} +1 \ge \sum_{i \le m} n_i^{6(B+1)} + 3m + 1$$
(since  each $n_i \ge 1$ and $m \ge 2$) 
$$ \ge \sum_{i \le m} P_{C_i}(n_i)+1$$
since each $C_i \le B$.
 It now remains to prove (5). 
We have $$ P_B(n_1 + \ldots + n_m ) = (n_1 + \ldots + n_m )^{6(B+1)} = \prod_{j \le m} (\sum_{i \le m} n_i)^{6(C_j+1)}$$
(since $\sum_{j \le m} (C_j +1) = B+1$)
$$ \ge \prod_{j \le m}((max_{i \le m} \, n_i +1)^2)^{3(C_j+1)} \ge \prod_{j \le m}(max_{i \le m} \, n_i^2 +2 +1)^{3(C_j+1)} $$
(since  each $n_i \ge 1$ and $m \ge 2$) 
$$    \ge \prod_{j \le m}(max_{i \le m} \, n_i^2 +2 )^{3(C_j+1)} +3  \ge \prod_{j \le m} (n_j^{6(C_j+1)}+ 2^3)  \ge \prod_{j \le m} P_B(n_j) +3,$$
 thus proving the Lemma.
\endproof

\begin{definition}\rm
Let $S$ be a schema. We define the set $W_S$ to be the subset of $ (\var \cup \f) \times \var$ for which both components occur in $S$. 
\end{definition}

\begin{definition}[recursive definition of $\datdep$ for a schema $S$] \label{datdep.defn}   \rm 

Let $S$ be a schema. Then $\datdep$ is the function $H$ from $W_S \times \vrtces(\gr(S)) $ to $\tf$ satisfying the following

\begin{enumerate}

\item

$H(v,v,\strt)= \tru$ for all  $(v,v) \in W_S$.

\item
 
If $w$ is a variable, $(l,l')$ is an edge in $\gr(S)$ and $\edgtyp(l,l')$ is not  an \ass\ to the variable $w$, then
 $H(f,w,l)= \tru \Ra   H(f,w,l')=\tru$ holds.
 
\item 

If $x,y \in \var$ and $(l,l')$ is an edge in $\gr(S)$ and $\edgtyp(l,l')$ is an \ass\ to the variable $y$
that references $x$, then $H(f,x,l)=\tru \Ra H(f,y,l)=\tru$ holds.
If in addition,  
the \ass\ $\assig(l,l')$ has function symbol $h$, then $H(h,y,l)=\tru$ holds,
\end{enumerate}
for which the set $H^{-1}(\tru)$ is minimal.
\end{definition}

\begin{theorem}\label{datdep.alg.equiv.thm}

Let $S$ be a free schema and let $(f,v) \in  W_S$. Let   \\ $l \in \vrtces(\gr(S))$. Then $\sed(f,v,l) \iff \datdep(f,v,l)$ holds.
 \end{theorem}

\proof 
Define the function $K: W_S \times \vrtces(\gr(S)) \to \tf$ as follows; $K(f,v,l)= \tru$ if and only if there is a  path $\mu$ through $S$ from $\strt $ to $l$ such that the term $\ee{\mu}{}(v)$ contains $f$. Since $S$ is free, $\sed = K $ holds. Thus it suffices to show that $K= \datdep$ holds, and this follows from the fact that Definition \ref{datdep.defn}, with $K$ in place of $\datdep$, gives an equivalent definition  of $K$. 
\endproof

The main Theorem of this subsection follows. 

\begin{theorem}\label{datdep.alg.poly.thm}
Let  $B\ge 0$ and let $S$ be a free schema in which every  $\conc$ construction occurs not more than $B$ times, and let $f \in \var \cup \f$ and $v \in \var$. Let $l \in \vrtces(\gr(S))$. Then it can be decided in polynomial time whether $\sed(f,v,l)$ holds.
\end{theorem}

\proof
From Theorem \ref{datdep.alg.equiv.thm} it suffices to prove that it can be decided in polynomial time whether $\datdep(f,v,l)$ holds, under the restriction given on $\conc$ constructions. We compute $\datdep(f,v,l)$ as follows, using the graph $\gr(S)$. \Wma\ that $(f,v) \in W_S$, since otherwise $\sed(f,v,l)$ can clearly be decided in polynomial time. 

We approximate  $\datdep$ on the domain $W_S \times \vrtces(\gr(S))$ by a sequence of functions  
$ H_1, H_2, \ldots : W \to \tf$. Firstly, let $H_1$ satisfy Condition (1) of Definition \ref{datdep.defn} for every $(v,v)\in W_S$ and let $ H_1(f,v,l)= \fal$ whenever $(f,v,l) \not= (v,v,\strt)$. Given a function $H_i$ that does not satisfy every instance of Condition (2) or (3) of Definition \ref{datdep.defn}, obtain the function $H_{i+1}$ by altering $H_i$ on one such  instance, so that  $H_{i+1}^{-1}(\tru)$ contains every element of  $H_{i}^{-1}(\tru)$, plus an additional one. Therefore a maximal function $H_n$ is eventually reached with $n \le W_S \times \vrtces(\gr(S))$, which is   polynomially bounded in terms of $S$, by 
 Lemma \ref{poly.time.graph.lem}. In addition, each function $H_i$ can be encoded by listing the elements of $H_{i}^{-1}(\tru)$, thus $H_n$ is computable in polynomial time.
 By induction on $i$, every set $H_{i}^{-1}(\tru) \subseteq \datdep^{-1}(\tru)$, and $H_n$ satisfies all three conditions in Definition \ref{datdep.defn}, hence the minimality condition in the definition of  $\datdep$ implies $H_n = \datdep$, thus proving the Theorem.
\endproof

\section{Conclusions}

We have extended conventional data dependency problems to arbitrary schemas and have shown that both the existential and universal data dependence problems lie in PSPACE for schemas without concurrency constructs and having only equality \asss,\ provided that there is a constant upper bound on the arity of any predicate symbol occurring in the schemas. We have also shown that without this upper bound, both problems are PSPACE-hard. This PSPACE-hardness result, Theorem \ref{hardness.pspace.sed.thm}, entails constructing a schema without this arity restriction; see the predicates $Q_3$ and $q_l$ in Figs. \ref{pspace.hard.fig} and  \ref{Tl.auto.fig}. This suggests that assuming this restriction 
 may result in a lower complexity bound than PSPACE. Since schemas with predicates approximate the behaviour of real programs much more accurately than wholly non-deterministic programs which are normally used in program analysis, a reasonable class of schemas for which our two problems could be decided tractably would be of considerable interest.

In addition, we have proved that for free schemas, existential data dependence is decidable in polynomial time provided that a constant upper  bound is placed on the number of occurrences of $\conc$ in the schemas being considered. We have not attempted to prove an analogous result for the universal data dependence relation. This would be an interesting subject for future investigation.

As mentioned in the Introduction, many concurrent systems have only relatively few threads even if they have many lines of code, and therefore the bound on the number of occurrences of $\conc$ is not particularly restrictive. The freeness hypothesis (equivalent to assuming non-deterministic branching) is common in program analysis, and its use  ensures that no  false positives for data dependence are computed.  This is important in areas such as security where we wish to show that the value of one variable $x$, whose value is accessible, cannot depend on the value of another variable $y$ whose value should be kept secret.

\textit{Acknowledgements} 
We would like to thank the referees for their helpful comments and detailed reviews of an earlier version of this paper.

\bibliographystyle{acmtrans}
\bibliography{slice240605}

\end{document}

%% file: commands.acm.tex
\newcommand{\new}{\newcommand} 
 \new{\np}{\newpage} 
 \new{\md}{\medskip} 
 \new{\sm}{\smallskip}
\new{\ra}{\rightarrow} 
\new{\la}{\leftarrow}
\new{\Ra}{\Rightarrow} 
\new{\La}{\Leftarrow}
 \new{\cen}{\center}

\new{\eg}{for example}
\new{\Txs}{There exists}
\new{\txs}{there exists}
\new{\tx}{there exist} 
\new{\Tx}{There exist}
\new{\ter}{interpretation}
\new{\ters}{interpretations}
\new{\st}{such that}
 \new{\ifa}{if and only if} 
 \new{\ass}{assignment} 
 \new{\asss}{assignments}
\new{\equ}{equivalent} 
\new{\Wma}{We may assume} 
\new{\wma}{we may assume}
\new{\wrt}{with respect to}
\new{\enum}{\begin{enumerate}}
\new{\enumE}{\end{enumerate}}
\new{\sat}{satisfy}
\new{\sating}{satisfying}
\new{\otoh}{on the other hand}
\new{\Otoh}{On the other hand}
\new{\ih}{induction hypothesis}
\new{\iha}{induction hypothesis applied}

\newtheorem{theorem}{Theorem}[section]

\newtheorem{lemma}[theorem]{Lemma}
\newdef{definition}[theorem]{Definition}
\newdef{remark}[theorem]{Remark}

\new{\as}{\,{\tt {:=}}\,}
\new{\whi}{\,\mathit{while}\,}
\new{\du}{\,\mathit{do}\,}
\new{\then}{\,\mathit{then}\,}
\new{\els}{\,\mathit{else}\,}
\new{\beginn}{\text{\it begin }}
\new{\eend}{\text{\it end }}
\new{\tru}{{\mathsf T}}
\new{\fal}{{\mathsf F}}
\new{\si}{\,\mathit{if}\,}
\new{\ski}{\mathit{skip}}
\new{\sch}{\mathit{Sch}}  
\new{\Int}{\mathit{Int}}
\new{\term}{\mathit{Term}} 
\new{\sig}{\mathit{Ass}}

\new{\p}{\mathcal{P}}
\new{\f}{\mathcal{F}}
\new{\labels}{{\mathcal L}}

 \new{\var}{\mathcal{V}} 
\new{\emp}{\varepsilon}

\new{\tl}{\triangleleft}
 \new{\fA}{\Gamma}
 \new{\fS}{S}
 \new{\fT}{T}
 \new{\lang}{{\mathit alphabet}}
 \new{\allpath}{\Pi^\omega}
 
 \new{\pathset}{\Pi}
 \new{\pathinf}[1]{\Pi^{\infty}(#1)} %not much used
 \new{\pathacm}[2]{{\pi}_{#1} ( #2)}        %schema, (int, state)
 
 \new{\set}[1]{\{ #1\}}
 
 \new{\subterms}{\mathit{Subterms}}
 \new{\termfun}{\mathit{TermSymbols}}
 \new{\pathlets}{\mathit{PathSymbols}}
 \new{\predpass}{\mathit{Passpred}}
 \new{\funpass}{\mathit{Passfunc}}
 
 \new{\terlets}[1]{{\mathit{InterLetters}}_{#1}}
 \new{\sub}[1]{\,{\mathit{sub}}_{#1} \,}
 \new{\subtrans}[1]{\,{\mathit{sub}^*}(#1) \,}
 
 \new{\state}{{\rm State}}
 \newcommand{\quine}[1]{ [ \!  [{#1}] \!  ] }
 
 \new{\arity}{\mathit{arity}}
 
 \new{\schema}{\mathit{schema}}
 
 \new{\predpath}[1]{\mathit{Predpaths}_{\, #1}}
\new{\func}{\mathit{Funcs}}
 \new{\pred}{\mathit{Preds}}
 \new{\ifpred}{\mathit{ifPreds}}
 \new{\whipred}{\mathit{whilePreds}}
 \new{\dd}[2]{{\mathcal M}\quine{#1}^{#2}_d}
 \new{\ee}[2]{{\mathcal M}\quine{#1}^{#2}_e}     %first arg is schema, second is int
  \new{\mean}[3]{{\mathcal M}{\quine{#1}^{#2}_{{#3}}}}  %schema,int,state
 \new{\supe}[3]{#1_{#2}^{#3}} 
 \new{\pre}{\mathit{pre}}
\new{\maxpre}{\mathit{maxpre}}
 
\newcommand\vx{{\mathbf{x}}}
\newcommand\vy{{\mathbf{y}}}

\new{\vt}{{\mathbf{t}}}

\newcommand{\vs}{\mathbf{s}}
\newcommand{\vv}{\mathbf{v}}
\newcommand{\vw}{\mathbf{w}}

\new{\ul}{\underline}

\newcommand{\perc}[1][S]{{\underset{#1}{\rightsquigarrow}}}
\new{\Perc}[1][S]{ !!\rightsquigarrow_{#1} }
\new{\finperc}[1][S]{\rightsquigarrow_{#1}^{\text{final}}}

\new{\cont}[1][S]{ \, \searrow_{#1} \, }
\new{\tcont}[1][S]{ \, {\searrow_{#1}}^{*} \; }

\new{\thru}[1][\fS]{\longmapsto_{#1}}
\new{\weak}[1]{\cong_{#1}}
\new{\strong}[1]{\mathbf{\cong_{#1}}}
\new{\seq}{\mathit{Sequences}}
\new{\mseq}{\mathit{maxSequences}}
\new{\isdefd}{\stackrel{\mathrm{def}}{=}}
\new{\assig}[1][S]{ \mathit{assign}_{#1} }
\new{\refset}[1][S]{ \mathit{refVars}_{#1} }
\new{\refvec}[1][S]{ \mathbf{refvec}_{#1} }

\new{\sli}[1][V]{ {{\mathit{sl}_{#1}}} }
\new{\inv}[1][S]{\mathit{Inv}_{#1}}
\new{\comps}{\mathit{Comps}}
\new{\opn}{ < \! \!\!}
\new{\clo}{\!\! >}
\new{\eqspace}{\!\! =\!\!}
\new{\simi}[1][R,V]{\tilde_{#1}}
\new{\need}[1][S]{{\mathcal N}_{{#1}}}
\new{\tword}[3]{\opn{#1}\supe{({#2})}{}{#3}\clo}
\new{\occ}{\mathit{Occ}}
\new{\gr}[1][S]{\mathit{grade}_{#1}}
\new{\grx}[1][S]{\mathit{grade}^{\mathit EXT}_{#1}}
\new{\grint}[1][S]{\mathit{grade}^{\mathit INT}_{#1}}
\new{\dex}[1][S]{\mathit{index}_{#1}}
\new{\dexx}[1][S]{\mathit{index}^{\mathit EXT}_{#1}}
\new{\inte}{\mathit{int}}
\new{\exte}{\mathit{EXT}}
\new{\seg}{\mathit{Seg}}
\new{\comb}{\bowtie}
\new{\letter}[2]{\opn {#1}\,{\clo}\!\!_{#2}\,}
\new{\pletter}[2]{\opn {#1},#2\,{\clo}}
\new{\body}[1][S]{ {{\mathit{body}_{#1}}} }
\new{\tf}{\{\tru,\fal\}}
\new{\op}{\neg}

\new{\outif}[1][S]{ {{\mathit{outif}_{#1}}} }
\new{\outwhi}[1][S]{ {{\mathit{outwhile}_{#1}}} }
\new{\out}[1][S]{ {{\mathit{out}_{#1}}} }
\new{\throu}[1][S]{ {{\mathit{thru}_{#1}}} }
\new{\back}[1][S]{ {{\mathit{back}_{#1}}} }
\new{\con}{\mathcal C}
\new{\varr}{\mathit whileVar}
\new{\corr}[1][S,T]{\mathit{Corr}_{#1}}
\new{\scorr}[1][S,T]{\mathit{StrongCorr}_{#1}}
\new{\precc}[1][S]{<\!\!<_{#1}}
\new{\abuv}[1][S]{\mathit{above}_{#1}}

\new{\tail}[1][S]{\mathit{tail}_{#1}}
\new{\head}[1][S]{\mathit{head}_{#1}}
\new{\tailterm}[1][S]{\mathit{Tailterms}_{#1}}
\new{\headterm}[1][S]{\mathit{Headterms}_{#1}}
\new{\com}[1][S]{\mathit{tail}_{#1}}

\new{\simbol}{\mathit{symbol}}
\new{\lsimbol}{\mathit{symbol}^{(\l)}}
\new{\sym}[1][S]{\mathit{Symbols}_{#1}}
\new{\lsym}[1][S]{\mathit{{Symbols}^{\mathcal L}}_{#1}}
\new{\ifprec}[1][p,S]{\sqsubseteq_{#1}}
\new{\pass}[1][S]{\mathit{passage}_{#1}}
\new{\Pass}{\mathit{Passages}}
\new{\init}{\mathit init}
\new{\next}[1][S]{{{\mathit{nextsymbol}}_{#1}}}

\new{\fir}{\mathit{firstpred}}
\new{\swhi}{\mathit{swhilePreds}}
\new{\ia}{immediately above}
\new{\sw}{super-while}

\new{\simil}[1][u]{\,\mathord{\mathit{simil}_{#1}\,}}
\new{\class}{\mathit{class}}
 \new{\trunc}[1][S]{\mathit{trunc}_{#1}\,}
 \new{\dep}[1][S]{\mathit{depnum}_{#1}\,}

 \new{\px}{p(\vx)}
 \new{\py}{p(\vy)}
 \new{\qx}{q(\vx)}
 \new{\qy}{q(\vy)}

 \new{\subb}{\subseteq}
  \new{\supp}{\supseteq}

  \new{\goto}{\mathit{goto\,\,}}

 \new{\ppair}[1][S]{\mathit{Predpairs}_{#1}}

\new{\cng}[1][u]{\mathit{cong}_{#1}\,}

\new{\simclass}[2][S]{[#2]_{#1}}

\new{\sto}{\mathit{STOP}}
\new{\start}{\mathit{START}}

\new{\before}[1][S]{<\!\!<_#1}

\new{\weis}[1][S]{{\mathcal W}_{#1}}
\new{\wpath}[1][S]{{\mathcal W}{\mathit paths}_{#1}}
\new{\wpred}[1][S]{{\mathcal W}{\mathit preds}_{#1}}
\new{\wfunc}[1][S]{{\mathcal W}{\mathit funcs}_{#1}}
\new{\wsym}[1][S]{{\mathcal W}{\mathit symbols}_{#1}}

\new{\lfnl}{LFn-L}
\new{\lib}{{\mathit libFuncs}}
\new{\cfunc}{{\mathit constFuncs}}

\def\labl{^{(l)}}

\new{\ls}[2][l]{{#2}^{(#1)}}
\new{\labs}{\mathit{labels}}          %label, then function symbol

 \new{\lfunc}{{\func^{\mathcal L}}}
\new{\lpred}{\pred^{\mathcal L}}
 \new{\lifpred}{{\ifpred^{\mathcal L}}}
 \new{\lwhipred}{{\whipred^{\mathcal L}}}

\new{\resp}{respectively}

\new{\gra}{\mathit Graph}
\new{\simcl}[2][S]{[#2]_{#1}}

 \new{\last}{\mathit{end}}
\new{\proj}[1][S']{ {\mathit{proj}}_{#1} }

\new{\scheq}{\text{:}\!\!\!=}

\new{\ubef}{\before[\text{}]}
\new{\usclass}[1]{[#1]}

\new{\call}{{\mathit call}}
\new{\uncall}{{\mathit uncall}}

\new{\conc}{{\parallel}}

\new{\sed}[1][S]{ \exists {\mathit{DD}}_{#1}    }

\new{\sad}[1][S]{ \forall {\mathit{DD}}_{#1}    }

\new{\strt}{\mathbf{start}}
\new{\nd}{\mathbf{end}}

\new{\emptwrd}{\varepsilon}

\new{\frst}{\mathit{firstlabel}}

\new{\datdep}[1][S]{\exists\mathit{DatDep}_{#1}}
\new{\datdap}[1][S]{\forall\mathit{DatDep}_{#1}}
\new{\vrtces}{\mathit{Vertices}}

\new{\lop}{\mathbf{loop\;}}